\documentclass{article}
\usepackage{amssymb}

\usepackage{latexsym}
\usepackage{amsmath}
\usepackage{graphicx}
\usepackage{float}

\newlength{\dinwidth}
\newlength{\dinmargin}
\begin{document}



\thispagestyle{empty} \vspace*{1cm} 
\vspace*{2cm}

\begin{center}
{\LARGE Paired quantum Hall states on noncommutative two-tori }

{\LARGE \ }

{\large Vincenzo Marotta\footnote{{\large {\footnotesize Dipartimento di
Scienze Fisiche,}\textit{\ {\footnotesize Universit\'{a} di Napoli
``Federico II''\ \newline
and INFN, Sezione di Napoli}, }{\small Compl.\ universitario M. Sant'Angelo,
Via Cinthia, 80126 Napoli, Italy}}},} {\large Adele Naddeo\footnote{{\large
{\footnotesize CNISM, Unit\`{a} di Ricerca di Salerno and Dipartimento di
Fisica \textit{''}E. R. Caianiello'',}\textit{\ {\footnotesize %
Universit\'{a} degli Studi di Salerno, }}{\small Via Salvador Allende, 84081
Baronissi (SA), Italy}}}}

{\small \ }

\textbf{Abstract\\[0pt]
}
\end{center}

\begin{quotation}
By exploiting the notion of Morita equivalence for field theories on
noncommutative tori and choosing rational values of the noncommutativity
parameter $\theta $ (in appropriate units), a one-to-one correspondence
between an abelian noncommutative field theory (NCFT) and a non-abelian
theory of twisted fields on ordinary space can be established. Starting from
this general result, we focus on the conformal field theory (CFT) describing
a quantum Hall fluid (QHF) at paired states fillings $\nu =\frac{m}{pm+2}$
\cite{cgm2}, recently obtained by means of $m$-reduction procedure, and show
that it is the Morita equivalent of a NCFT. In this way we extend the
construction proposed in \cite{AV1} for the Jain series $\nu =\frac{m}{2pm+1}
$. The case $m=2$ is explicitly discussed and the role of noncommutativity
in the physics of quantum Hall bilayers is emphasized. Our results represent
a step forward the construction of a new effective low energy description of
certain condensed matter phenomena and help to clarify the relationship
between noncommutativity and quantum Hall fluids.

\vspace*{0.5cm}


{\footnotesize PACS: 11.25.Hf, 11.10.Nx, 73.43.Cd\newpage }\baselineskip%
=18pt \setcounter{page}{2}
\end{quotation}

\section{Introduction}

The quantum Hall effect (QHE) is one of the most remarkable many-body
phenomena discovered in the last twenty-five years \cite{prangegirvin}\cite
{perspectives}. It takes place in a two-dimensional electron gas formed in a
quantum well in a semiconductor host material and in the presence of a very
high magnetic field \cite{dorda}\cite{stormer}, as a result of the
commensuration between the number of electrons $N_{e}$ and the number of
flux quanta $N_{\Phi }$. The electrons condense into distinct and highly
non-trivial ground states (`vacua') formed at each integer (IQHE) \cite
{laughlin1} or rational fractional value (FQHE) \cite{laughlin2} of the
filling factor $\nu =\frac{N_{e}}{N_{\Phi }}$. In particular at fractional
fillings quasi-particles with fractional charge and statistics emerge, and
new kinds of order parameters are considered \cite{ODLRO}. In such a context
strong research interests have been growing in the last years towards a full
understanding of the physics of those plateaux which do not fall into the
hierarchical scheme \cite{jain}. To such an extent a pairing picture, in
which pairs of spinless or spin-polarized fermions condense, has been
proposed \cite{MR} for the non-standard fillings $\nu =\frac{1}{q}$, $q>0$
and even. As a result the ground state has been described in terms of the
Pfaffian (the so called Pfaffian state) and the non-Abelian statistics of
the fractional charged excitations evidenced \cite{MR}\cite{FNTW}. Today
quasi-particles with non-Abelian statistics appear very promising in view of
the realization of a fault-tolerant quantum computer \cite{fault1}. Indeed
protection against decoherence could be obtained by encoding quantum
information in some topological characteristics of the strongly correlated
electron system while quantum gates could be implemented by topologically
non trivial operations such as braidings of non-Abelian anyons.

Following this line, increasing technological progress in molecular beam
epitaxy techniques has led to the ability to produce pairs of closely spaced
two-dimensional electron gases. Since then such bilayer quantum Hall systems
have been widely investigated theoretically as well as experimentally \cite
{perspectives}\cite{teoria, eisenstein}. Strong correlations between the
electrons in different layers lead to new physical phenomena involving
spontaneous interlayer phase coherence with an associated Goldstone mode. In
particular a spontaneously broken $U(1)$ symmetry \cite{Zee} has been
discovered and identified and many interesting properties of such systems
have been studied: the Kosterlitz-Thouless transition, the zero resistivity
in the counter-current flow, a DC/AC Josephson-like effect in interlayer
tunneling as well as the presence of a gapless superfluid mode \cite{Zee1}
\cite{girvin1}. Indeed, when tunneling between the layers is weak, the
quantum Hall bilayer state can be viewed as arising from the condensation of
an excitonic superfluid in which an electron in one layer is paired with a
hole in the other layer. The uncertainty principle makes it impossible to
tell which layer either component of this composite boson is in.
Equivalently the system may be regarded as a ferromagnet in which each
electron exists in a coherent superposition of the ''pseudospin''
eigenstates, which encode the layer degrees of freedom \cite{YangMoon}\cite
{girvin1}. The phase variable of such a superposition fixes the orientation
of the pseudospin magnetic moment and its spatial variations govern the low
energy excitations in the system. So quantum Hall bilayers are an
interesting realization of the pairing picture at non-standard fillings.

On the other hand, noncommutative field theories (NCFT) have attracted much
attention in the last years because they provide a non trivial
generalization of local quantum field theories, allowing for some degree of
non locality while retaining an interesting mathematical structure \cite
{ncft1}. One motivation for the relevance of such theories is that the
notion of space-time presumably has to be modified at very short distances
by introducing a limit to the resolution in which one may probe it. In this
way space-time becomes fuzzy at very short distances because the concept of
a \textit{point} is lost.

The simplest framework in which NCFT emerge in a natural way as an effective
description of the dynamics is just the Landau problem \cite{landau1},
namely the quantum mechanics of the motion of $N_{e}$ charged particles in
two dimensions subjected to a transverse magnetic field. The strong field
limit $B\rightarrow \infty $ at fixed mass $m$ projects the system onto the
lowest Landau level and, for each particle $I=1,...,N_{e}$, the
corresponding coordinates ($\frac{eB}{c}x_{I},y_{I}$) are a pair of
canonical variables which satisfy the commutation relations $\left[
x_{I},y_{I}\right] =i\delta _{I,J}\theta $, $\theta =\frac{\hbar c}{eB}%
\equiv l_{M}^{2}$ being the noncommutativity parameter. In this picture the
electron is not a point-like particle and can be localized at best at the
scale of the magnetic length $l_{M}$.

Indeed an alternative description is possible, which reminds a system of
open strings in a $B$-field background \cite{ncft3}, the one in terms of
neutral fields such as the density and the current which carry the quantum
numbers of dipoles. As a consequence, the currents and the density of a
system of electrons in a strong magnetic field may be described by a
noncommutative Chern-Simons theory \cite{qhf1}, which captures the granular
structure of the fluid and exactly reproduces the quantitative connection
between filling fractions and statistics, a crucial feature of quantum Hall
fluid (QHF) physics \cite{qh1}\cite{top1}. An equivalent formulation is
possible as well in terms of a matrix theory similar to that describing $D0$%
-branes in string theory \cite{brane1}. These observations stimulated a new
research area, so in the last years deep efforts have been devoted to a
whole understanding of the relationship between noncommutative spaces and
QHF and its implications. As a result, noncommutativity is related to the
finite number $N_{e}$ of electrons in a realistic sample via the rational
parameter $\theta \propto \frac{1}{N_{e}}$, which sets the elementary area
of nonlocality \cite{qhf2}\cite{qhf3}\cite{qhf4}. In this context the $%
\theta $-dependence of physical quantities is expected to be analytic near $%
\theta =0$ \cite{alvarez1} because of the very smooth ultraviolet behavior
of noncommutative Chern-Simons theories \cite{uv1}; as a consequence the
effects of the electron's granularity embodied in $\theta $ can be expanded
in powers of $\theta $ via the Seiberg-Witten map \cite{ncft2} and then
appear as corrections to the large-$N$ results of field theory.

A significant feature of NCFT is the Morita duality between noncommutative
tori \cite{morita}\cite{scw1}\cite{scw2}, which establishes a relation, via
a one-to-one correspondence, between representations of two noncommutative
algebras. If we refer to gauge theories on noncommutative tori, Morita
duality can be viewed as a low energy analogue of $T$-duality of the
underlying string model \cite{string1} and, as such, it results a powerful
tool in order to establish a correspondence between NCFT and well known
standard field theories. Indeed, for rational values of the noncommutativity
parameter, $\theta =\frac{1}{N}$, one of the theories obtained by using the
Morita equivalence is a commutative field theory of matrix valued fields
with twisted boundary conditions and magnetic flux $c$ \cite{wilson1}
(which, in a string description, behaves as a $B$-field modulus). Also open
Wilson lines of the noncommutative theory are mapped to closed Wilson lines
wrapping the torus of the commutative theory \cite{matrix1}. The $U\left(
N\right) $ commutative theory displays a bit unexpected behavior, for
instance in correspondence of particular values of the relevant parameters
its renormalized dispersion relation will develop tachyonic modes \cite
{guralnik1} which appear to be related to the spontaneous $Z_{N}\times Z_{N}$
symmetry breaking due to electric flux condensation. In other words, the
commutative field theory keeps track of the spontaneous breaking of
translational invariance which characterizes its dual (noncommutative)
counterpart.

In a recent work \cite{AV1} we followed such an approach focusing on a
particular conformal field theory (CFT), the one obtained via $m$-reduction
technique \cite{VM}, which has been successfully applied to the description
of a quantum Hall fluid (QHF) at Jain \cite{cgm1}\cite{cgm3} as well as
paired states fillings \cite{cgm2}\cite{cgm4} and in the presence of
topological defects \cite{noi1}\cite{noi2}\cite{noi5}. In particular, we
showed by means of the Morita equivalence that a NCFT with $\theta =2p+\frac{%
1}{m}$ is mapped to a CFT on an ordinary space. We identified such a CFT
with the $m$-reduced CFT developed for a QHF at Jain fillings $\nu =\frac{m}{%
2pm+1}$ \cite{cgm1}\cite{cgm3}, whose neutral fields satisfy twisted
boundary conditions. In this way we gave a meaning to the concept of
''noncommutative conformal field theory'', as the Morita equivalent version
of a CFT defined on an ordinary space. The image of Morita duality in the
ordinary space is given by the $m$-reduction technique and the corresponding
noncommutative torus Lie algebra is naturally realized in terms of
Generalized Magnetic Translations (GMT). That introduces a new relationship
between noncommutative spaces and QHF and paves the way for further
investigations on the role of noncommutativity in the physics of general
strongly correlated many body systems \cite{ncmanybody}.

In this work, by making use of Morita duality, we will show that an abelian
NCFT with $\theta =\frac{p}{2}+\frac{1}{m}$ is mapped to a nonabelian theory
of twisted fields on ordinary space, which coincides with the $m$-reduced
CFT developed for a QHF at paired states fillings $\nu =\frac{m}{pm+2}$ \cite
{cgm2}\cite{cgm4}. That extends to non standard fillings the analysis
carried out in Ref. \cite{AV1}. In particular we focus on the $m=2$ case
which is experimentally relevant and describes a system of two parallel
layers of $2D$ electrons gas in a strong perpendicular magnetic field. The
consequences of noncommutativity on the physics of the pairing picture in
quantum Hall bilayers are analyzed in detail.

The paper is organized as follows.

In Section 2, we review the description of a QHF at paired states fillings $%
\nu =\frac{m}{pm+2}$ obtained by means of the $m$-reduction procedure \cite
{cgm2,cgm4}. We focus mainly on the $m=2$ case and briefly recall the
physics of quantum Hall bilayers on the plane.

In Section 3, we explicitly build up the Morita equivalence between CFTs in
correspondence of rational values of the noncommutativity parameter $\theta $
with an explicit reference to the $m$-reduced theory describing a QHF at
paired states fillings. Indeed we clearly show that there is a well defined
one-to-one correspondence between the fields on a noncommutative torus and
those of a non-abelian field theory on an ordinary space.

In Section 4, we show how the noncommutative torus Lie algebra is realized
through GMT in the QHF context. The role of the noncommutative tori in the
twisted sector of our theory is evidenced and the relevance of non-Abelian
statistics of the corresponding ground states is outlined. Finally, the
implications on the transport properties of a quantum Hall bilayer with
different boundary conditions are briefly discussed.

In Section 5, some comments and outlooks of this work are given.

Finally, in the Appendix we recall the $m$-reduction description of a QHF at
paired states fillings on the torus topology, focusing on the special $m=2$
case \cite{cgm4}.

\section{The $m$-reduction description of a QHF at paired states fillings}

In this Section we review how the $m$-reduction procedure on the plane
(genus $g=0$) \cite{VM} works in describing successfully a QHF at paired
states fillings $\nu =\frac{m}{pm+2}$ \cite{cgm2,cgm4}. We focus mainly on
the special case $m=2$ and on the physics of a quantum Hall bilayer.

The idea is to build up an unifying theory for all the plateaux with even
denominator starting from the bosonic Laughlin filling $\nu =1/pm+2$, which
is described by a CFT (mother theory) with $c=1$, in terms of a scalar
chiral field compactified on a circle with radius $R^{2}=1/\nu =pm+2$ (or
the dual $R^{2}=4/pm+2$). Then the $U(1)$ current is given by $%
J(z)=i\partial _{z}Q(z)$, where $Q(z)$ is the compactified Fubini field with
the standard mode expansion:
\begin{equation}
Q(z)=q-i\,p\,lnz+\sum_{n\neq 0}\frac{a_{n}}{n}z^{-n},  \label{modes}
\end{equation}
with $a_{n}$, $q$ and $p$ satisfying the commutation relations $\left[
a_{n},a_{n^{\prime }}\right] =n\delta _{n,n^{\prime }}$ and $\left[ q,p%
\right] =i$. Let us notice that the informations about the quantization of
momentum and the winding numbers are stored in the lattice geometry induced
by the QHE quantization (see Ref. \cite{cgm2} for details); in other words
the QHE physics fixes the compactification radius. The corresponding primary
fields are expressed in terms of the vertex operators $U^{\alpha
}(z)=:e^{i\alpha Q(z)}:$ with $\alpha ^{2}=1,...,2+pm$ and conformal
dimension $h=\frac{\alpha ^{2}}{2}$.

Starting with this set of fields and using the $m$-reduction procedure,
which consists in considering the subalgebra generated only by the modes in
Eq. (\ref{modes}), which are a multiple of an integer $m$, we get the image
of the twisted sector of a $c=m$ orbifold CFT (daughter theory), the twisted
model (TM), which describes the lowest Landau level dynamics. Then the
fields in the mother CFT can be factorized into irreducible orbits of the
discrete $Z_{m}$ group which is a symmetry of the daughter theory and can be
organized into components which have well defined transformation properties
under this group. The general characteristics of the daughter theory is the
presence of twisted boundary conditions (TBC) which are induced on the
component fields and are the signature of an interaction with a localized
topological defect \cite{noi1,noi2,noi5}. When we generalize the
construction to a torus (genus $g=1$), we find different sectors
corresponding to different boundary conditions imposed at the ends of the
finite two-dimensional ($2D$) layer, as shown in detail in Refs. \cite{cgm4}
\cite{cgm3} and briefly recalled in the Appendix. To compare the orbifold so
built with the $c=m$ CFT, we use the mapping $z\rightarrow z^{1/m}$and the
isomorphism defined in Ref.\cite{VM} between fields on the $z$ plane and
fields on the $z^{m}$ covering plane given by the following identifications:
$a_{nm+l}\longrightarrow \sqrt{m}a_{n+l/m}$, $q\longrightarrow \frac{1}{%
\sqrt{m}}q$.

Let us now focus on the special $m=2$ case, which describes a system
consisting of two parallel layers of $2D$ electrons gas in a strong
perpendicular magnetic field. The filling factor $\nu ^{(a)}=\frac{1}{2p+2}$
is the same for the two $a=1$, $2$ layers while the total filling is $\nu
=\nu ^{(1)}+\nu ^{(2)}=\frac{1}{p+1}$. For $p=0$ ($p=1$) it describes the
bosonic $220$ (fermionic $331$) Halperin (H) state \cite{Halperin}.

The CFT description for such a system can be given in terms of two
compactified chiral bosons $Q^{(a)}$ with central charge $c=2$. In order to
construct the fields $Q^{(a)}$ for the TM, the starting point is the bosonic
filling $\nu =1/2(p+1)$, described by a CFT with $c=1$ in terms of a scalar
chiral field $Q$ compactified on a circle with radius $R^{2}=1/\nu =2(p+1)$
(or its dual $R^{2}=2/(p+1)$), see Eq. (\ref{modes}). The $m$-reduction
procedure generates a daughter theory which is a $c=2$ orbifold. Its primary
fields content can be expressed in terms of a $Z_{2}$-invariant scalar field
$X(z)$, given by
\begin{equation}
X(z)=\frac{1}{2}\left( Q^{(1)}(z)+Q^{(2)}(-z)\right) ,  \label{X1}
\end{equation}
describing the electrically charged sector of the new filling, and a twisted
field
\begin{equation}
\phi (z)=\frac{1}{2}\left( Q^{(1)}(z)-Q^{(2)}(-z)\right) ,  \label{phi1}
\end{equation}
which satisfies the twisted boundary conditions $\phi (e^{i\pi }z)=-\phi (z)$
and describes the neutral sector \cite{cgm2}. Such TBC signal the presence
of a localized topological defect which couples, in general, the $m$ edges
in a $m$-layers system \cite{noi1}\cite{noi2}\cite{noi5}. In the case of our
interest, a bilayer system ($m=2$), we get a crossing between the two edges
as sketched in Fig. 1.

\begin{figure}[ht]
\centering\includegraphics*[width=0.8\linewidth]{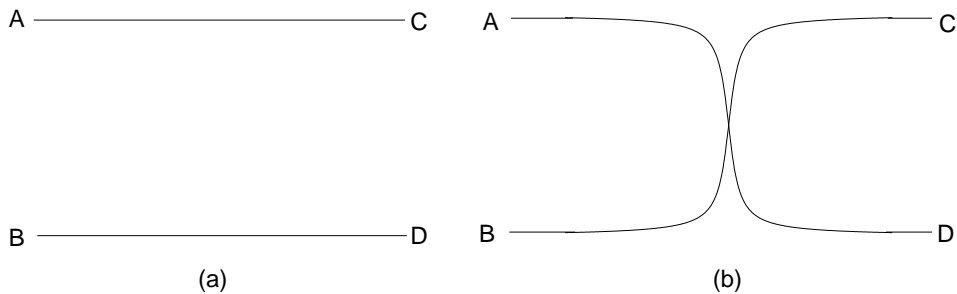}
\caption{The bilayer system, (a) without the topological defect (PBC), (b)
with the topological defect (TBC).}
\label{figura2}
\end{figure}

The chiral fields $Q^{(a)}$, defined on a single layer $a=1$, $2$, due to
the boundary conditions imposed upon them by the orbifold construction, can
be thought of as components of a unique ``boson''\ defined on a double
covering of the disc (layer) ($z_{i}^{(1)}=-z_{i}^{(2)}=z_{i}$). As a
consequence the two layers system becomes equivalent to one-layer QHF (in
contrast with the Halperin model in which they appear independent) and the $%
X $ and $\phi $ fields defined in Eqs. (\ref{X1}) and (\ref{phi1})
diagonalize the interlayer interaction. In particular the $X$ field carries
the total charge with velocity $v_{X}$, while $\phi $ carries the charge
difference of the two edges with velocity $v_{\phi }$, i.e. no charge, being
the number of electrons the same for each layer (balanced system).

The primary fields are the composite operators $V(z)=\mathcal{U}_{X}(z)\psi
(z)$, where $\mathcal{U}_{X}(z)=\frac{1}{\sqrt{z}}:e^{i\alpha X(z)}:$ are
the vertices of the charged sector with $\alpha ^{2}=2(p+1)$. Furthermore
the highest weight states of the neutral sector can be classified in terms
of two kinds of chiral operators, $\psi (z)\left( \bar{\psi}(z)\right) =%
\frac{1}{2\sqrt{z}}\left( e^{i\alpha {\cdot }\phi (z)}\pm ie^{i\alpha {\cdot
}\phi (-z)}\right) $, which, in a fermionic language, correspond to $c=1/2$
Majorana fermions with periodic (Ramond) or anti-periodic (Neveu-Schwarz)
boundary conditions \cite{cgm4}. As a consequence this theory decomposes
into a tensor product of two CFTs, a twisted invariant one with $c=3/2$,
realized by the charged boson $X(z)$ and the Ramond Majorana fermion,\ which
is coupled to the charged sector, while the second one has $c=1/2$ and is
realized in terms of the Neveu-Schwarz Majorana fermion. The two Majorana
fermions just defined are inequivalent, due to the breaking of the symmetry
which exchanges them and that results in a non-Abelian statistics. Such a
factorization results much more evident in the construction of the modular
invariant partition function, as we briefly recall in the Appendix \cite
{cgm4}. The bosonized energy-momentum tensor of the $Z_{2}$ twist invariant
theory develops a cosine term in its neutral sector which is described by
the Ramond fields:
\begin{equation}
T_{\psi _{s}}(z)=-\frac{1}{4}(\partial \phi )^{2}-\frac{1}{16z^{2}}\cos (2%
\sqrt{2}\phi ).
\end{equation}
It is a clear signature of a tunneling phenomenon which selects out a new
stable vacuum,\ the $c=3/2$ one. If we refer to the bilayer system, we can
reduce the spacing between the layers so that the two species of electrons
which live on them become indistinguishable: in such a case the tunneling
amplitude gets large enough to make the H states flow to the Moore-Read (MR)
states \cite{MR}. In the limit of strong tunneling the velocity of one
Majorana becomes zero and the theory reduces to the $c=3/2$ CFT. Let us also
point out that $m$-ality in the neutral sector is coupled to the charged one
exactly, according to the physical request of locality of the electrons with
respect to the edge excitations. Indeed our projection, when applied to a
local field, automatically couples the discrete $Z_{m}$ charge of $U(1)$
with the neutral sector in order to give rise to a single valued composite
field.

Now let us give an interpretation of the existence of these sectors in terms
of conformal invariant boundary conditions which are due to the scattering
of the particles on localized impurities \cite{noi1}\cite{noi2}\cite{noi5}.
The H sector describes a pure QHF phase in which \ no impurities are present
and the two layers edges are not connected (see Fig. 1(a)). In realistic
samples however this is not the case and the deviations from the Halperin
state may be regarded as due to the presence of localized impurities. These
effects can be accounted for by allowing for more general boundary
conditions just as the ones provided by our TM. In fact an impurity located
at a given point on the edge induces twisted boundary conditions for the
boson $\phi $ and, as a consequence, a current can flow between the layers.
Then a coherent superposition of interlayer interactions could drive the
bilayer to a more symmetric phase in which the two layers are
indistinguishable due to the presence of a one electron tunneling effect
along the edge.

The primary fields content of the theory just introduced on the torus
topology will be given in the Appendix.

\section{$m$-reduced CFT for QHF at paired states fillings and Morita
equivalence}

In this Section we exploit the notion of Morita equivalence on
noncommutative tori for rational values of the noncommutativity parameter $%
\theta $ and construct a general one-to-one correspondence between NCFTs and
CFTs on the ordinary space. We will refer to the $m$-reduced theory
describing a QHF at paired states fillings, recalled in Section 2. In this
way we extend our previous results concerning a QHF at Jain hierarchical
fillings \cite{AV1} and further clarify the role of noncommutativity in QHF
physics by introducing a new perspective.

The Morita equivalence \cite{morita}\cite{scw1}\cite{scw2} is defined in
full generality as a one-to-one correspondence between representations of
two noncommutative algebras; indeed it conserves all the modules and their
associated structures. Let us now consider an $U(N)$ NCFT defined on the
noncommutative torus $\mathrm{T}_{\theta }^{2}$ and, for simplicity, of
radii $R$. Such a model has $N\times N$ matrix degrees of freedom and arises
from the regularization of the theory describing a membrane with
world-volume topology $\mathrm{T}^{2}\times \mathrm{R}$. In general, this
procedure is related to the fact that the symmetry group of area-preserving
diffeomorphisms on the membrane can be approximated by $U(N)$ for a surface
of any genus \cite{matrix2}\cite{membrane1}; furthermore it is equivalent to
approximating the membrane surface by a finite lattice.

The simplest noncommutative space is $\mathrm{R}_{\theta }^{2}$ with
coordinates satisfying the Heisenberg commutation relation $%
[x_{1},x_{2}]=i\Theta $. Along with the space $\mathrm{R}_{\theta }^{2}$ it
is natural to consider its compactification $\mathrm{T}_{\theta }^{2}$ \cite
{ncft1} with the Morita duality represented by the following $SL(2,Z)$
action on the parameters:
\begin{equation}
\theta ^{^{\prime }}=\frac{a\theta +b}{c\theta +d};\text{ \ \ \ \ \ \ \ \ }%
R^{^{\prime }}=\left| c\theta +d\right| R;  \label{morita1}
\end{equation}
here $\theta =\Theta /\left( 2\pi R^{2}\right) $\textrm{, }$a,b,c,d$ are
integers and $ad-bc=1$. An intriguing feature arises for rational values of
the non commutativity parameter, $\theta =-\frac{b}{a}$, when $c\theta +d=%
\frac{1}{a}$: the Morita transformation (\ref{morita1}) sends the NCFT to an
ordinary one with $\theta ^{^{\prime }}=0$ and different radius $R^{^{\prime
}}=\frac{R}{a}$, involving in particular a rescaling of the rank of the
gauge group \cite{scw1}\cite{scw2}\cite{alvarez1}. Indeed the dual theory is
a twisted $U(N^{^{\prime }})$ theory with $N^{^{\prime }}=aN$ and 't Hooft
flux $c$. The classes of $\theta ^{^{\prime }}=0$ theories are parametrized
by an integer $m$, so that for any $m$ there is a finite number of abelian
theories which are related by a subset of the transformations given in Eq. (%
\ref{morita1}). Our main result relies on such consideration and can be
expressed as follows.

The $m$-reduction technique applied to the QHF at paired states fillings ($%
\nu =\frac{m}{pm+2}$, $p$ even) can be viewed as the image of the Morita map
(characterized by $a=\frac{p}{2}\left( m-1\right) +1$, $b=\frac{p}{2}$, $%
c=m-1$, $d=1$) between the two NCFTs with $\theta =1$ and $\theta =\frac{p}{2%
}+\frac{1}{m}$ ($\theta =\nu _{0}/\nu ,$ being $\nu _{0}=1/2$ the filling of
the starting theory), respectively and corresponds to the Morita map in the
ordinary space. The $\theta =1$ theory is an $U\left( 1\right) _{\theta =1}$
NCFT while the mother CFT is an ordinary $U\left( 1\right) $ theory;
furthermore, when the $U\left( 1\right) _{\theta =\frac{p}{2}+\frac{1}{m}}$
NCFT is considered, its Morita dual CFT has $U\left( m\right) $ symmetry. As
a consequence, the following correspondence Table between the NCFTs and the
ordinary CFTs is established:
\begin{equation}
\begin{array}{ccc}
& \text{Morita} &  \\
U\left( 1\right) _{\theta =1} & \rightarrow & U\left( 1\right) _{\theta =0}
\\
& \left( a=1,b=-1,c=0,d=1\right) &  \\
\text{Morita}\downarrow \left( a,b,c,d\right) &  & m-\text{reduction}%
\downarrow \\
& \text{Morita} &  \\
U\left( 1\right) _{\theta =\frac{p}{2}+\frac{1}{m}} & \rightarrow & U\left(
m\right) _{\theta =0} \\
& \left( a=m,b=-\frac{pm}{2}-1,c=1-m,d=\frac{p}{2}\left( m-1\right) +1\right)
&
\end{array}
\label{morita2}
\end{equation}
Let us notice that theories which differ by an integer in the
noncommutativity parameter are not identical because they differ from the
point of view of the CFT. In fact, the Morita map acts on more than one
parameter of the theory. For instance, the compactification radius of the
charged component is renormalized to $R_{X}^{2}=p+\frac{2}{m}$,\ that gives
rise to different CFTs by varying $p$ values. Moreover the action of the $m$%
-reduction procedure on the number $p$ doesn't change the central charge of
the CFT under study but modifies the spectrum of the charged sector \cite
{cgm2}\cite{cgm4}.

In order to show that the $m$-reduction technique applied to the QHF at
paired states fillings is the image of the Morita map between the two NCFTs
with $\theta =1$ and $\theta =\frac{p}{2}+\frac{1}{m}$ respectively and
corresponds to the Morita map in the ordinary space it is enough to show how
the twisted boundary conditions on the neutral fields of the $m$-reduced
theory (see Section 2) arise as a consequence of the noncommutative nature
of the $U\left( 1\right) _{\theta =\frac{p}{2}+\frac{1}{m}}$ NCFT.

Let us start by recalling that an associative algebra of smooth functions
over the noncommutative two-torus $\mathrm{T}_{\theta }^{2}$ ($\left[
x_{1},x_{2}\right] =i\Theta $) can be realized through the Moyal star
product:
\begin{equation}
f\left( x\right) \ast g\left( x\right) =\left. \exp \left( \frac{i\Theta }{2}%
\left( \partial _{x_{1}}\partial _{y_{2}}-\partial _{x_{2}}\partial
_{y_{1}}\right) \right) f\left( x\right) g\left( y\right) \right\vert _{y=x}.
\label{moyal1}
\end{equation}
Such functions are identified with general field operators $\Phi $ and, when
defined on a torus, can have different boundary conditions associated to any
of the compact directions. For the torus we have four different
possibilities:
\begin{equation}
\begin{array}{cc}
\Phi \left( x_{1}+R,x_{2}\right) =e^{2\pi i\alpha _{1}}\Phi \left(
x_{1},x_{2}\right) , & \Phi \left( x_{1},x_{2}+R\right) =e^{2\pi i\alpha
_{2}}\Phi \left( x_{1},x_{2}\right) ,
\end{array}
\label{bc1}
\end{equation}
where $\alpha _{1}$ and $\alpha _{2}$ are the boundary parameters. It is
useful to decompose the elements of the algebra in their Fourier components.
The Fourier expansion of the general field operator $\Phi _{\overrightarrow{%
\alpha }}$ with boundary conditions $\overrightarrow{\alpha }=\left( \alpha
_{1},\alpha _{2}\right) $ takes the form:
\begin{equation}
\Phi _{\overrightarrow{\alpha }}=\sum_{\overrightarrow{n}}\Phi ^{%
\overrightarrow{n}}U_{\overrightarrow{n}+\overrightarrow{\alpha }},
\label{fexp1}
\end{equation}
where we define the generators of the algebra as
\begin{equation}
U_{\overrightarrow{n}+\overrightarrow{\alpha }}\equiv \exp \left( 2\pi i%
\frac{\left( \overrightarrow{n}+\overrightarrow{\alpha }\right) \cdot
\overrightarrow{x}}{R}\right) .  \label{fexp2}
\end{equation}
They give rise to the following commutator:
\begin{equation}
\left[ U_{\overrightarrow{n}+\overrightarrow{\alpha }},U_{\overrightarrow{%
n^{\prime }}+\overrightarrow{\alpha ^{\prime }}}\right] =-2i\sin \left( 2\pi
\theta \left( \overrightarrow{n}+\overrightarrow{\alpha }\right) \wedge
\left( \overrightarrow{n^{\prime }}+\overrightarrow{\alpha ^{\prime }}%
\right) \right) U_{\overrightarrow{n}+\overrightarrow{n^{\prime }}+%
\overrightarrow{\alpha }+\overrightarrow{\alpha ^{\prime }}},  \label{fexp3}
\end{equation}
where $\overrightarrow{p}\wedge \overrightarrow{q}=\varepsilon
_{ij}p_{i}q_{j}$.

When the noncommutativity parameter $\theta $ takes a rational value, which
we choose as $\theta =\frac{q}{m}$, being $q=\frac{p}{2}m+1$ and $m$
relatively prime integers, the infinite-dimensional algebra generated by the
$U_{\overrightarrow{n}+\overrightarrow{\alpha }}$ breaks up into equivalence
classes of finite dimensional ($m\times m$) subspaces. Indeed the elements $%
U_{m\overrightarrow{n}}$ generate a center of the algebra and, within Eq. (%
\ref{fexp1}), can be treated as ordinary generators defined on a commutative
space. That makes possible for the momenta the following decomposition:
\begin{equation}
\overrightarrow{n^{\prime }}+\overrightarrow{\alpha }=m\overrightarrow{n}+%
\overrightarrow{j},\text{ \ \ \ \ }0\leq j_{1},j_{2}\leq m-1.  \label{fexp4}
\end{equation}
and leads to the splitting of the whole algebra into equivalence classes
classified by all the possible values of $m\overrightarrow{n}$. Each class
is a subalgebra generated by the $m^{2}$ functions $U_{\overrightarrow{n}+%
\overrightarrow{j}}$ which satisfy the relations
\begin{equation}
\left[ U_{\overrightarrow{n}+\overrightarrow{j}},U_{\overrightarrow{%
n^{\prime }}+\overrightarrow{j^{\prime }}}\right] =-2i\sin \left( \frac{2\pi
q}{m}\overrightarrow{j}\wedge \overrightarrow{j^{\prime }}\right) U_{%
\overrightarrow{n}+\overrightarrow{n^{\prime }}+\overrightarrow{j}+%
\overrightarrow{j^{\prime }}}.  \label{fexp5}
\end{equation}
The algebra (\ref{fexp5}) just introduced is isomorphic to the
(complexification of the) $U\left( m\right) $ algebra, whose general $m$%
-dimensional matrix representation can be constructed by means of the
following ''shift''\ and ''clock''\ generators \cite{matrix1}\cite{matrix2}
\cite{matrix3}:
\begin{equation}
Q=\left(
\begin{array}{cccc}
1 &  &  &  \\
& \varepsilon  &  &  \\
&  & \ddots  &  \\
&  &  & \varepsilon ^{m-1}
\end{array}
\right) ,\text{ \ \ \ \ \ \ }P=\left(
\begin{array}{cccc}
0 & 1 &  & 0 \\
& \cdots  &  &  \\
&  & \vdots  & 1 \\
1 &  &  & 0
\end{array}
\right) ,  \label{fexp6}
\end{equation}
being $\varepsilon =\exp (\frac{2\pi i}{m})$. Matrices $P$ and $Q$ are
unitary, traceless and satisfy the property:
\begin{equation}
PQ=\varepsilon QP.  \label{weyl}
\end{equation}
So the matrices $J_{\overrightarrow{j}}=\varepsilon ^{\frac{j_{1}j_{2}}{2}%
}Q^{j_{1}}P^{j_{2}}$, $j_{1},j_{2}=0,...,m-1$, generate an algebra
isomorphic to (\ref{fexp5}):
\begin{equation}
\left[ J_{\overrightarrow{j}},J_{\overrightarrow{j^{\prime }}}\right]
=-2i\sin \left( 2\pi \frac{q}{m}\overrightarrow{j}\wedge \overrightarrow{%
j^{\prime }}\right) J_{\overrightarrow{j}+\overrightarrow{j^{\prime }}}.
\label{fexp7}
\end{equation}
Thus the following Morita mapping has been realized between the Fourier
modes defined on a noncommutative torus and $U\left( m\right) $-valued
functions defined on a commutative space:
\begin{equation}
\exp \left( 2\pi i\frac{\left( \overrightarrow{n}+\overrightarrow{j}\right)
\cdot \widehat{\overrightarrow{x}}}{R}\right) \longleftrightarrow \exp
\left( 2\pi i\frac{\overrightarrow{n}\cdot \overrightarrow{x}}{R}\right) J_{%
\overrightarrow{j}}.  \label{fexp8}
\end{equation}
As a consequence a mapping between the general field operator $\mathbf{\Phi }
$ on the noncommutative torus \textrm{T}$_{\theta }^{2}$ and the field $\Phi
$ on the dual commutative torus $\mathrm{T}_{\theta =0}^{2}$ is generated as
follows:
\begin{equation}
\Phi =\sum_{\overrightarrow{n}}\exp \left( 2\pi im\frac{\overrightarrow{n}%
\cdot \widehat{\overrightarrow{x}}}{R}\right) \sum_{\overrightarrow{j}%
=0}^{m-1}\Phi ^{\overrightarrow{n},\overrightarrow{j}}U_{\overrightarrow{n}+%
\overrightarrow{j}}\longleftrightarrow \Phi =\sum_{\overrightarrow{j}%
=0}^{m-1}\chi ^{\left( \overrightarrow{j}\right) }J_{\overrightarrow{j}}.
\label{fexp9}
\end{equation}
The new field $\Phi $ is defined on the dual torus with radius $R^{\prime }=%
\frac{R}{m}$ and satisfies the \textit{twist eaters} boundary conditions,
due to the presence of the 't Hooft magnetic flux $c$:
\begin{equation}
\begin{array}{cc}
\Phi \left( x_{1}+R^{\prime },x_{2}\right) =\Omega _{1}^{+}\cdot \Phi \left(
x_{1},x_{2}\right) \cdot \Omega _{1}, & \Phi \left( x_{1},x_{2}+R^{\prime
}\right) =\Omega _{2}^{+}\cdot \Phi \left( x_{1},x_{2}\right) \cdot \Omega
_{2},
\end{array}
\label{fexp12}
\end{equation}
with
\begin{equation}
\Omega _{1}=P^{c},\text{ \ \ \ \ \ \ }\Omega _{2}=Q,  \label{fexp13}
\end{equation}
where $c$ is an integer satisfying $dq-cm=1$. Furthermore the field
components $\chi ^{\left( \overrightarrow{j}\right) }$, defined as:
\begin{equation}
\chi ^{\left( \overrightarrow{j}\right) }=\exp \left( 2\pi i\frac{%
\overrightarrow{j}\cdot \overrightarrow{x}}{R}\right) \sum_{\overrightarrow{n%
}}\Phi ^{\overrightarrow{n},\overrightarrow{j}}\exp \left( 2\pi im\frac{%
\overrightarrow{n}\cdot \overrightarrow{x}}{R}\right) ,  \label{fexp11}
\end{equation}
satisfy the following twisted boundary conditions:
\begin{equation}
\begin{array}{c}
\chi ^{\left( \overrightarrow{j}\right) }\left( x_{1}+R^{\prime
},x_{2}\right) =e^{2\pi ij_{1}/m}\chi ^{\left( \overrightarrow{j}\right)
}\left( x_{1},x_{2}\right)  \\
\chi ^{\left( \overrightarrow{j}\right) }\left( x_{1},x_{2}+R^{\prime
}\right) =e^{2\pi ij_{2}/m}\chi ^{\left( \overrightarrow{j}\right) }\left(
x_{1},x_{2}\right)
\end{array}
,  \label{fexp14}
\end{equation}
that is
\begin{equation}
\left( \frac{j_{1}}{m},\frac{j_{2}}{m}\right) ,\text{ \ \ \ \ }%
j_{1}=0,...,m-1,\text{ \ \ \ \ }j_{2}=0,...,m-1.  \label{fexp15}
\end{equation}
Within a gauge theory context, generally modes with fractional momentum $%
\overrightarrow{j}$ carry the electric flux:
\begin{equation}
e_{i}\equiv q\epsilon _{il}j_{l}modm.  \label{electric1}
\end{equation}

Let us notice that $\chi ^{\left( 0,0\right) }$ is the trace degree of
freedom which can be identified with the $U(1)$ component of the matrix
valued field or the charged $X$ field within the $m$-reduced theory of the
QHF at paired states fillings introduced in Section 2, while the twisted
fields $\chi ^{\left( \overrightarrow{j}\right) }$ with $\overrightarrow{j}%
\neq \left( 0,0\right) $ should be identified with the neutral ones (\ref
{phi1}). The commutative torus is smaller by a factor $m\times m$ than the
noncommutative one; in fact upon this rescaling also the \textquotedblright
density of degrees of freedom\textquotedblright\ is kept constant as now we
are dealing with $m\times m$ matrices instead of scalars. In conclusion,
when the parameter $\theta $ is rational we recover the whole structure of
the noncommutative torus and recognize the twisted boundary conditions which
characterize the neutral fields (\ref{phi1}) of the $m$-reduced theory as
the consequence of the Morita mapping of the starting NCFT ($U\left(
1\right) _{\theta =\frac{p}{2}+\frac{1}{m}}$ in our case) on the ordinary
commutative space.

The key role in the proof of equivalence is played by the map on the field $%
Q(z)$\ of Eq. (\ref{modes}) which, after the Morita action, is defined on
the noncommutative space $z\rightarrow z^{1/m}\equiv U_{0,1}$. The algebra
defined by the commutation rules in Eq. (\ref{fexp5}) is realized in terms
of the $m^{2}-1$ general operators:
\begin{equation}
\begin{array}{cc}
U_{j_{1},j_{2}}=\varepsilon ^{\frac{j_{1}j_{2}}{2}}z^{j_{1}}\varepsilon
^{j_{2}\widetilde{\sigma }}, &
\begin{array}{c}
j_{1},j_{2}=0,...,m-1 \\
\left( j_{1},j_{2}\right) \neq \left( 0,0\right)
\end{array}
\end{array}
,  \label{circles4}
\end{equation}
\bigskip where $\widetilde{\sigma }=iz\partial _{z}$. By using the
decomposition given in Eq. (\ref{fexp9}), where $\Phi \equiv Q(z)$, we
identify the fields $X(z)$\ and $\phi ^{j}(z)$\ of the CFT defined on the
ordinary space. At this point the correspondence becomes explicit due to the
equivalence between the correlators evaluated on the noncommutative space
(by using the operators $a_{nm+l}\longrightarrow \sqrt{m}a_{n+l/m}$, $%
q\longrightarrow \sqrt{\theta }q$) and the corresponding ones obtained for
the commutative case. The correlators of the neutral component $\phi ^{j}(z)$
become identical to those given by a set of $m-1$\ bosonic fields with
twisted boundary conditions (see Refs. \cite{cgm2}\cite{cgm4} for details).

From the CFTs point of view, two theories are equivalent if they have the
same symmetry (i.e. Virasoro or Kac-Moody algebras, for instance) and the
same correlators on the plane and on the torus (i.e. the conformal blocks).
Therefore the NCFT built of the field $Q$\ on $\mathrm{T}_{\theta }^{2}$,
whose symmetry, correlators and conformal blocks were calculated in
accordance in \cite{cgm2}\cite{VM}, is equivalent to the CFT composed of the
charged field $X$\ and the $m-1$\ twisted fields $\phi ^{j}(z)$\ with the same
symmetry, correlators and conformal blocks (see Refs. \cite{VM}\cite{cgm4}).

In the following Section we further clarify such a correspondence by making
an explicit reference to the QHF physics at paired states fillings. In
particular we will recognize the GMT as a realization of the noncommutative
torus Lie algebra defined in Eq. (\ref{fexp5}).

\section{Generalized magnetic translations and noncommutative torus Lie
algebra}

In this Section we construct GMT on a torus within the $m$-reduced theory
for a QHF at paired states fillings introduced in Section 2 and show that
they are a realization of the noncommutative torus Lie algebra. That
completes the proof of our main claim (see Section 3): the $m$-reduction map
is the counterpart of Morita map in the ordinary space and the $m$-reduced
theory keeps track of noncommutativity in its structure. Then we realize
that such a structure is shared also by a pure Yang-Mills theory. An
emphasis on the particular $m=2$ case is given, which corresponds to a
quantum Hall bilayer with different boundary conditions \cite{noi1}\cite
{noi2}\cite{noi5}. The action of GMT on the characters of our TM is given
together with the corresponding physical interpretation. In general, the
different possible boundary conditions are associated with different
possible impurities (topological defects). In turn, a boundary state can be
defined in correspondence to each class of defects \cite{noi1} and
identified with a condensate of t'Hooft electric and magnetic fluxes. In
this context the GMT can be identified with the Wilson loop operators and
act on the boundary states, that is electric-magnetic condensates \cite
{AVfuture1}. In the language of Kondo effect \cite{kondo1} they behave as
boundary condition changing operators. We will report in detail on such
issues in a forthcoming publication \cite{AVfuture1}.

It is today well known that QHF are quantum liquids with novel and extremely
rich internal structures, the so called topological orders \cite{wen}.
Topological order is recognized as a kind of order which cannot be obtained
by means of a spontaneous symmetry breakdown mechanism and, as such, it is
responsible of all the unusual properties of QHF. In this context the role
of translational invariance is crucial in determining the relevant
topological effects, such as the degeneracy of the ground state wave
function on a manifold with non trivial topology, the derivation of the Hall
conductance $\sigma _{H}$ as a topological invariant and the relation
between fractional charge and statistics of anyon excitations \cite
{cristofano1}. All the above phenomenology strongly relies on the invariance
properties of the wave functions under a finite subgroup of the magnetic
translation group for a $N_{e}$ electrons system. Indeed their explicit
expression as the Verlinde operators \cite{top3}, which generate the modular
transformations in the $c=1$ CFT, are taken as a realization of topological
order of the system under study \cite{wen}. In particular the magnetic
translations built so far \cite{cristofano1} act on the characters
associated to the highest weight states which represent the charged
statistical particles, the anyons or the electrons.

In our CFT representation of the QHF at paired states fillings \cite{cgm2}
\cite{cgm4} we shall see that the noncommutative torus Lie algebra defined
in Eq. (\ref{fexp5}) has a natural and beautiful realization in terms of
GMT. We refer to them as generalized ones because the usual magnetic
translations act on the charged content of the one point functions \cite
{cristofano1}. Instead, in our TM model for the QHF (see Section 2, Appendix
and Refs. \cite{cgm3}, \cite{cgm4}) the primary fields (and then the
corresponding characters within the torus topology) appear as composite
field operators which factorize in a charged as well as a neutral part.
Further they are also coupled by the discrete symmetry group $Z_{m}$. Then,
in order to show that the characters of the theory are closed under magnetic
translations, we need to generalize them in such a way that they will appear
as operators with two factors, acting on the charged and on the neutral
sector respectively. The presence of the transverse magnetic field $B$
reduces the torus to a noncommutative one and the flux quantization induces
rational values of the noncommutativity parameter $\theta $.

Let us consider a general magnetic translation of step $\left(
n=n_{\upharpoonright }+in_{\downharpoonleft },\overline{n}%
=n_{\upharpoonright }-in_{\downharpoonleft }\right) $ on a sample with
coordinates $\left( x_{1},x_{2}\right) $ and denote with $T^{n,\overline{n}}$
the corresponding generators. Let us denote with $\upharpoonleft $ and $%
\downharpoonright $ the layer index because we are dealing with a bilayer
system. Within our TM for a QHF at paired states fillings \cite{cgm2}\cite
{cgm4} it is possible to show that such generators can be factorized into a
group which acts only on the charged sector as well as a group acting only
on the neutral sector \cite{AVfuture2}. In this context the $\mathit{%
classical}$ magnetic translations group considered in the literature
corresponds to the TM charged sector. In order to study the action of a GMT
on the torus and clarify its interpretation in terms of noncommutative torus
Lie algebra, Eq. (\ref{fexp7}), let us evaluate how the argument of the
Theta functions in which the conformal blocks are expressed gets modified.
For a bilayer Hall system a translation carried out on the layer $%
\upharpoonleft $ or $\downharpoonright $ produces a shift in the layer Theta
argument $w_{i}$, $w_{i}\rightarrow w_{i}+\delta _{i}$, which can be
conveniently expressed in terms of the charged and neutral ones $w_{c(n)}=%
\frac{w_{\upharpoonleft }\pm w_{\downharpoonright }}{2}$, and in this way we
obtain the action on the conformal blocks of the TM. Indeed, from the
periodicity of the Theta functions it is easy to show that the steps of the
charged and neutral translation can be parametrized by $\delta _{c}=\frac{%
2(p+1)l+2s+i}{2(p+1)}$ and $\delta _{n}=\pm l\pm \frac{i}{2}$ respectively,
being $l=0,1$; $s=0,...,p$; and $i=0,1$. The layer exchange is realized by
the transformation $w_{n}\rightarrow -w_{n}$ but the TM is built in such a
way to correspond to the exactly balanced system in which $w_{n}=0$\ (modulo
periodicity) so that this operation can be obtained only by exchanging the
sign in $\delta _{n}$ (independently for $l$ and $i$). Because of the
factorization of the effective CFT at paired states fillings into two
sub-theories with $c=3/2$ and $c=1/2$, corresponding to the MR and Ising
model respectively (see Section 2, Appendix and Refs. \cite{cgm2}\cite{cgm4}%
), we find that also GMT exhibit the same factorization \cite{AVfuture2}. As
a consequence, conformal blocks of MR and Ising sectors are stable under the
transport of electrons and of the neutral Ising fermion.

In order to derive the transformation properties of the TM characters we
resort to the following identity:
\begin{equation}
J_{c}^{\alpha ,\beta }K_{a}\left( w|\tau \right) =K_{a}\left( w+\frac{\alpha
\tau +\beta }{2(p+1)}|\tau \right) =e^{-2\pi i\frac{a\beta }{4(p+1)}%
}e^{-4\pi i\frac{\alpha \beta }{\left( 4(p+1)\right) ^{2}}}e^{-\pi i\frac{%
\alpha ^{2}\tau +4\alpha w}{4(p+1)}}K_{a+\alpha }\left( w|\tau \right)
\end{equation}
(the explicit expressions for $K_{a}\left( w|\tau \right) $ are given in the
Appendix). Let us notice that the factor in the above equation must be a
phase, so that in order to extend such a formula to any complex $\tau $ and $%
w$ we need to multiply the characters by a non-analytic factor, as
introduced in Ref. \cite{Cappelli}. Upon fixing $\mathbf{a}=2(p+1)\mathbf{l}+%
\mathbf{q}$ and $\mathbf{b}=2(p+1)\mathbf{l}^{\prime }+\mathbf{q}^{\prime }$%
, where $\mathbf{q}=2\mathbf{s}+\mathbf{i}$, and defining $\mathbf{a}{\times
}\mathbf{b}=a_{1}$ $b_{2}-a_{2}$ $b_{1}$ we obtain:
\begin{equation}
\left[ J_{c}^{\mathbf{a}},J_{c}^{\mathbf{b}}\right] =-2isin\left( 2\pi (%
\frac{p}{2}\mathbf{l}{\times }\mathbf{l}^{\prime }+\frac{\mathbf{l}{\times }%
\mathbf{l}^{\prime }}{2}+\frac{\mathbf{l}{\times }\mathbf{i}^{\prime }+%
\mathbf{i}{\times }\mathbf{l}^{\prime }}{4}+\frac{\mathbf{q}{\times }\mathbf{%
q}^{\prime }}{8(p+1)})\right) J_{c}^{\mathbf{a}+\mathbf{b}}.  \label{tmc1}
\end{equation}
For $p$ even, the first term does not give any contribution while different
interpretations can be given for the three kinds of excitations. For $\
i,i^{\prime }=0$ the characters are a $2(p+1)$-dimensional representation of
the Abelian magnetic translations generated by the transport of electrons ($%
s,s^{\prime }=0$) and anyons ($s,s^{\prime }\neq 0$) \ respectively. The $l$
($l^{\prime }$ resp.) index is the $Z_{2}$ charge of the parity rule and\
the coupling between $l$ and $i$, for $i,i^{\prime }\neq 0$, implies
non-Abelian statistics for the quasi-holes.

The $J_{c}^{0,\beta }$ translation, which acts as $w\rightarrow w+$ $\frac{%
\beta }{2(p+1)}$, can be viewed as the result of an electric potential $V$
while the $J_{c}^{\alpha ,0}$ translation, whose action is realized by $%
w\rightarrow w+$ $\frac{\alpha \tau }{2(p+1)}$, corresponds physically to
increase the flux through the sample, $N_{\Phi }=\Phi /\Phi _{0}$, by $\frac{%
\alpha }{2(p+1)}$ units of the elementary flux quantum $\Phi _{0}=hc/e$ and
that gives rise to the Laughlin spectral flow.

The charged characters $K_{\alpha }(w_{c}|\tau )$ appearing in Eqs. (\ref{MR}%
)-(\ref{MR2}) are closed under this group but the whole TM conformal blocks
are not stable. Nevertheless the translations (\ref{tmc1}) don't correspond
to physical ones due to the lack\ of the neutral sector contribution; so we
need to add the neutral sector in order to keep track of the layer degree of
freedom, i. e. the pseudospin. The magnetic translation of the whole
fundamental physical particles, i.e. the electrons with layer index,
corresponds to a translation of the charged as well as the neutral component
of the TM characters and is given by the shifts $\delta _{c},\delta
_{n}=(1,1)$ in the corresponding Theta function arguments. In this way we
find that the characters of the $331$ model, Eqs. (\ref{hal1})-(\ref{hal4}),
(which belong to the untwisted sector of our TM, see Appendix and Ref. \cite
{cgm4}) are invariants only under combined translations: $\chi
_{(a,s)}^{331}\rightarrow \chi _{(a,s)}^{331}$ for $a=1,..,4$. The exchange
of the last two characters is the signal of a layer exchange. Indeed such a
feature is irrelevant within the $331$ model because for a balanced system ($%
w_{n}=0$) the characters \ $\chi _{(3,s)}^{331}$, $\chi _{(4,s)}^{331}$ are
equivalent. Nevertheless it gives rise to interesting consequences within
the Ho model \cite{ho}, which takes into account the symmetrization of the
state with respect to the layer index. In this case there is only one state $%
\chi _{(3,s)}^{331}+\chi _{(4,s)}^{331}$, which forces the model to lie in
the balanced configuration. Within our TM model the balancing implies that
the bilayer system develops a defect, which gives rise to a contact point
between the two layers so allowing the transport of an electron from a layer
to another. That is the signature of the presence of a twist field in the
ground state. \ The factorization of TM into MR \ model and Ising one
implies that the two independent Ising models of the neutral sector are not
equivalent because the charged and the neutral sector of the MR model are
not completely independent but need to satisfy the constraint $\alpha \cdot
p+l=0$ (mod $2$) which is the $m$-ality condition (parity rule). It is
explicitly realized by constraining the eigenvalues of the fermion parity
operator, which is obtained by defining the generalized GSO projector as $P=%
\frac{1}{2}(1-e^{i\pi \alpha \cdot p}\gamma _{F})$. Here the operator $%
\gamma _{F}=(-1)^{F}$ is defined in such a way to anticommute with the
fermion field, $\gamma _{F}\psi \gamma _{F}=-\psi $, and to satisfy $\left(
\gamma _{F}\right) ^{2}=1$ ($F$ being the fermion number). It has
eigenvalues $\pm 1$ when acts on states with an even or an odd number of
fermion creation operators. In terms of GMT \ the above constraint implies
that charged and neutral translation cannot be independent.

Let us now observe that the emerging noncommutative torus introduces $m^{2}$
states but only $m$ are the diagonal ones. Due to the traceless condition
the neutral massless states are only $m-1$, so that there is only one state
in the bilayer case $m=2$. The natural interpretation of this scenario is
that $m(m-1)$ states are massive and the massless states are selected by the
projection operator $P$ above defined. That is also needed in order to
fulfill the consistence requirements imposed by modular invariance. The
degeneracy of the TM ground state implies the spontaneous breaking of the
fermion parity because the selected ground state has not a well defined
parity fermion number and non-Abelian statistics. In order to gain a
physical insight on the generalized GSO projection just described we point
out that there is a correspondence between TM vacua and particular
configurations of the bilayer Hall system \cite{noi1}\cite{noi2}. In this
way the twisted ground state corresponds to an exactly balanced system, i.e.
the layers are completely equivalent and there is no current flow by a layer
to another one. Finally, in order to formally extend the definition of
magnetic translations in the neutral sector to the transport of electrons
(or anyons) in any vacuum we have to introduce a couple ($a,F$) of
parameters which are defined only modulo $2$. Here $a$ is equal to $1$ or $0$
for twisted and untwisted vacua respectively. Such a definition is
consistent with the fusion rules of Ising model. In this way the transport
of a particle around another one produces an extra phase $(-1)^{\alpha
_{1}F_{2}-\alpha _{2}F_{1}}$, which completes the Aharonov--Bohm phase
arising from the charged sector.

From the above discussion it follows that the classification of excitations
in terms of \ $U(1)$ quantum numbers $l$, $s$, $i$ is misleading because the
TM characters $\chi _{(i,s)}^{\pm }$ are invariant under a greater algebra.
Therefore the GMT for TM\ can be decomposed into a subalgebra generating the
translations of the $l$-electrons, which is a symmetry of the model, and a
spectrum generating algebra which is written in terms of the $s$ index, the
spin $i$ and the $Z_{2}$ charge $\pm $ quantum numbers. The non-Abelian
nature of the MR model is the consequence of the lack of information on the
layer localization of the excitations. Indeed the pseudospin fusion rules
can be deduced from its interpretation in terms of layer index because, for
any $i=1$, two configurations exist with an excitation on the layer up $%
\upharpoonright $ or down $\downharpoonleft $. The MR model depends only on
the absolute value of the $\Sigma _{z}$ pseudospin. Therefore the state with
$l=0$ corresponds to the states ($\upharpoonleft \downharpoonleft \pm
\downharpoonleft \upharpoonleft $) while the states with $l=1$ are
identified with the states ($\upharpoonleft \upharpoonleft $ or $%
\downharpoonleft \downharpoonleft $). The degeneration can be resolved only
within the TM by means of the addition of the $\overline{Ising}$ quantum
numbers, which give us information about the localization of the excitation
in the terms of layers (i. e. the sign of the pseudospin).

The GMT operators above introduced realize an algebra isomorphic to the
noncommutative torus Lie algebra given in Eqs. (\ref{fexp5}) and (\ref{fexp7}%
). Such operators generate the residual symmetry of the $m$-reduced CFT
which is Morita equivalent to the NCFT with rational non commutativity
parameter $\theta =\frac{p}{2}+\frac{1}{m}$.

Let us now close this Section by making some interesting observations. The
GMT operators can be put in correspondence with the $SU\left( m\right) $
valued twist matrices associated to the cycles of the two-torus, $J_{a,0}$
and $J_{0,b}$, which satisfy the relation:
\begin{equation}
J_{a,0}J_{0,b}=J_{b,0}J_{0,a}e^{-2\pi iab/m}.  \label{gauge4}
\end{equation}
and may be chosen to be constant. In general, translations of the gauge
fields along the two cycles of the torus are equivalent to gauge
transformations \cite{matrix3}:
\begin{equation}
A_{i}\left( x+R_{j}\right) =U_{j}\left( x\right) A_{i}\left( x\right)
U_{j}^{+}\left( x\right) +U_{j}\left( x\right) i\partial _{i}U_{j}^{+}\left(
x\right) ,  \label{gauge1}
\end{equation}
where $R_{j}=R$, $j=1,2$, are the periodicities of the torus and we put $%
U_{1}=J_{a,0}$ and $U_{2}=J_{0,b}$. For adjoint matter the following
constraint holds on:
\begin{equation}
U_{i}\left( x\right) U_{j}\left( x+R_{i}\right) U_{i}^{+}\left(
x+R_{j}\right) U_{j}^{+}\left( x\right) =e^{2\pi ic_{ij}/m},  \label{gauge2}
\end{equation}
where the twist $c_{ij}$ is an integer and defines non-Abelian t'Hooft
magnetic fluxes through the non trivial cycles of the torus. If we consider
a Yang-Mills theory on $\mathrm{T}^{2}\times \mathrm{R}^{n}$ and choose the
time-like direction in $\mathrm{R}^{n}$, then in the $A_{0}=0$ gauge the
following twisted boundary conditions
\begin{equation}
U_{1}=P^{c},\text{ \ \ \ \ \ \ }U_{2}=Q,  \label{gauge3}
\end{equation}
are imposed, which act as:
\begin{equation}
A_{i}\left( x+R_{j}\right) =U_{j}\left( x\right) A_{i}\left( x\right)
U_{j}^{+}\left( x\right) .  \label{gauge1a}
\end{equation}
The choice (\ref{gauge3}) corresponds to the so called \textit{twist eaters}
matrices which generate the Weyl-t'Hooft algebra \cite{matrix3}. Let us
notice that different choices of twisted boundary conditions with the same
magnetic flux are related by large gauge transformations, but in the $%
A_{0}=0 $ gauge no paths exist, which connect different twisted boundary
conditions at different times. There exist other large gauge transformations
which leave the boundary conditions invariant, and their eigenvalues
determine the t'Hooft electric fluxes.

In this way the structure of GMTs within the $m$-reduced CFT describing the
physics of a QHF at paired states fillings coincides with that of a general $%
U\left( m\right) $ Yang-Mills gauge theory on a twisted two-torus. Indeed
noncommutativity makes the $U\left( 1\right) $ and $SU\left( m\right) $
sectors of the decomposition:
\begin{equation}
U\left( m\right) =U\left( 1\right) \times \frac{SU\left( m\right) }{Z_{m}}
\label{weyl2}
\end{equation}
not decoupled because the $U\left( 1\right) $ \textit{photon} interacts with
the $SU\left( m\right) $ \textit{gluons }\cite{gauge1}. That reflects the
nontrivial coupling between the different topological sectors in the $m$%
-reduced theory, or the interplay between charged and neutral degrees of
freedom in the QHF picture. In this sense the $m$-reduced theory retains
some degree of noncommutativity while the $m$-reduction map is the image of
the Morita map in the ordinary space.

The Wilson loop operators of Yang-Mills theory could be identified as GMT in
QHF context while the t'Hooft electric and magnetic flux condensates could
be identified with boundary states. In this way some features of
noncommutative gauge theories are recovered in condensed matter systems and
that has been recognized to be relevant, for instance, in the context of
topological quantum computation \cite{fault1}. Furthermore the interaction
in string and $D$-brane theory is introduced by means of gauge theories; one
could think that interactions between topological defects and QHF are
described by the same gauge theories. In order to clarify this analogy, let
us refer for simplicity to the $m=2$ case, corresponding to a quantum Hall
bilayer. Recently such a system has been shown to be well described within a
boundary CFT formalism and the possible defects (impurities) which are
compatible with conformal invariance as well as their stability (boundary
entropy \cite{ludaff1}) have been studied in detail \cite{noi1}\cite{noi2}
\cite{noi5}. To each class of defects corresponds a different boundary
state. For the bilayer system two possible boundary conditions are
identified, as shown in Fig. 1, the periodic (PBC) and twisted (TBC)
boundary conditions respectively, which give rise to different topological
sectors (vacua) on the torus. The defects break the GMT symmetry so that
different backgrounds can be connected by special GMT. The breaking of the
residual symmetry of the CFT can be recognized in the condensation of
electric-magnetic fluxes represented by the defects. The \textit{quarks} in
this picture appear as domain walls (or defects) interpolating between
different vacua and may carry fractional quantum numbers which differ from
those of the electron. The coupling of two QHF layers at the boundary is a
topological one and it is well described \cite{noi2} by a boundary magnetic
term of the kind \cite{callan1}:
\begin{equation}
S_{mag}=i\frac{\beta }{4\pi }\int_{0}^{T}dt\left( X\partial _{t}Y-Y\partial
_{t}X\right) _{\sigma =0},  \label{boundarymag1}
\end{equation}
where $X$ and $Y$ are two massless scalar fields in $1+1$ dimensions and $%
\beta =2\pi B$ is related to the magnetic field orthogonal to the $X-Y$
plane. If we resort to a string analogy, such a term allows for exchange of
momentum of the open string moving in an external magnetic field. Indeed
within the worldsheet field theory for open strings attached to $D$-branes,
it can be written as \cite{magneticbrane1}:
\begin{equation}
S_{\partial \Sigma }=-\frac{i}{2}\oint_{\partial \Sigma }dtB_{ij}y^{i}\left(
t\right) \overset{.}{y}^{j}\left( t\right) ,  \label{boundarymag2}
\end{equation}
where $t$ is the coordinate of the boundary $\partial \Sigma $ of the string
worldsheet residing on the $D$-brane worldvolume, $B_{ij}$ is a magnetic
field on the $D$-branes and $y^{i}\left( t\right) $ are the open string
endpoint coordinates. Notice that Eq. (\ref{boundarymag2}) formally
coincides with that of the Landau problem in the strong field limit and then
canonical quantization of the coordinates $y^{i}\left( t\right) $ will
induce a noncommutative geometry on the $D$-brane worldvolume. In this way
noncommutative field theories emerge as effective descriptions of the string
dynamics in a background $B$-field. The deep relation between the boundary
CFT given by Eq. (\ref{boundarymag1}) and NCFT, as derived by its string
counterpart (\ref{boundarymag2}), will be fully exploited within the system
of two QHF layers analyzed here in a forthcoming publication \cite{AVfuture1}%
.

\section{Conclusions and outlooks}

This work relies strongly on the power of Morita equivalence, which allows
one to establish a one-to-one correspondence between representations of two
noncommutative algebras, and on a peculiar feature of such a correspondence:
if the starting field theory is defined on a noncommutative torus and the
noncommutativity parameter $\theta $ is rational, the dual theory is a
commutative field theory of matrix valued fields with twisted boundary
conditions and magnetic flux $c$ \cite{scw1}\cite{scw2}. In this paper, by
extending a recent proposal \cite{AV1}, we have shown by means of the Morita
equivalence that a NCFT with $\theta =\frac{p}{2}+\frac{1}{m}$ is mapped to
a CFT on an ordinary space. We identified such a CFT with the $m$-reduced
CFT developed in \cite{cgm2}\cite{cgm4} for a QHF at paired states fillings,
whose neutral fields satisfy twisted boundary conditions. The $m$-reduction
technique appears as the image in the ordinary space of the Morita duality
and, as such, retains some noncommutative features: the non trivial
interplay between charged and neutral degrees of freedom as well as the
structure of GMT, which realize the noncommutative torus Lie algebra in the
QHF context. In this way a picture emerges on the interplay between
noncommutativity and QHF physics, which is very different from the ones
developed in the literature in the last years \cite{qhf1}\cite{qhf2}\cite
{qhf3}\cite{qhf4}\cite{ncmanybody}. The identification of objects of pure
Yang-Mills theory such as Wilson loop operators and t'Hooft electric and
magnetic flux condensates with GMT and boundary states respectively in the
context of a QHF bilayer system will be carried out in detail in a
forthcoming publication together with a careful analysis of the relation
between the corresponding boundary CFT and NCFT provided by string theory
\cite{AVfuture1}.

Recently, the twisted CFT approach provided by $m$-reduction and developed
for a QHF at paired states fillings \cite{cgm2}\cite{cgm4} has been
successfully applied to Josephson junction ladders and arrays of non trivial
geometry in order to investigate the existence of topological order and
magnetic flux fractionalization in view of the implementation of a possible
solid state qubit protected from decoherence \cite{noi3,noi4,noi6} as well
as to the study of the phase diagrams of the fully frustrated $XY$ model ($%
FFXY$) on a square lattice \cite{noi} and of a general spin-1/2
antiferromagnetic two leg ladder with Mobius boundary conditions in the
presence of various perturbations \cite{noi7}. So, it could be interesting
to generalize our approach in order to further elucidate the topological
properties of Josephson systems with non trivial geometry as well as the
deep nature of the quantum critical points and of the deconfinement of
excitations with fractional quantum numbers in antiferromagnetic spin
ladders and the consequences imposed upon them by the request of space-time
noncommutativity. That could open new perspectives on the general relation
between noncommutativity and the physics of strongly correlated electron
systems.

On the other hand, the $m$-reduction technique has been recently employed in
order to shed new light on the analogy between string theory and QHF physics
\cite{branehall2}. By generalizing the quantum Hall soliton introduced in
Ref. \cite{branehall1} the tachyon condensation in a non-BPS system of $D$%
-branes \cite{tachyon1} has been found similar to the tunneling between two
layers of QHF at paired states fillings. In that context the role of the $%
Z_{2}$ ($Z_{m}$) symmetry has been pointed out: indeed such a discrete
symmetry couples the charged and the neutral vertices hinting to a more
general description in terms of Matrix Theory with $U(2)$ ($U(m)$) symmetry
\cite{brane1}. Following the line introduced in this work, Morita duality
could help us to shed new light on the connections between a system of
interacting $D$-branes and the physics of quantum Hall fluids through the
unifying framework of Matrix String Theory and CFT \cite{AVfuture}.

\section*{Appendix: $m$-reduction procedure for a QHF at paired states
fillings on the torus}

In this Appendix we give the whole primary fields content on the torus
topology of the theory introduced in Section 2. We focus on the particular
case $m=2$, which describes the physics of a quantum Hall bilayer.

On the torus, the primary fields are described in terms of the conformal
blocks of the MR and the Ising model \cite{cgm4}. The MR characters $\chi
_{(\lambda ,s)}^{MR}$ with $\lambda =0,...,2$ and $s=0,...,p$, are
explicitly given by:
\begin{eqnarray}
\chi _{(0,s)}^{MR}(w|\tau ) &=&\chi _{0}(\tau )K_{2s}\left( w|\tau \right)
+\chi _{\frac{1}{2}}(\tau )K_{2(p+s)+2}\left( w|\tau \right)  \label{MR} \\
\chi _{(1,s)}^{MR}(w|\tau ) &=&\chi _{\frac{1}{16}}(\tau )\left(
K_{2s+1}\left( w|\tau \right) +K_{2(p+s)+3}\left( w|\tau \right) \right)
\label{MR1} \\
\chi _{(2,s)}^{MR}(w|\tau ) &=&\chi _{\frac{1}{2}}(\tau )K_{2s}\left( w|\tau
\right) +\chi _{0}(\tau )K_{2(p+s)+2}\left( w|\tau \right) .  \label{MR2}
\end{eqnarray}
They represent the field content of the $Z_{2}$ invariant $c=3/2$ CFT \cite
{MR} with a charged component ($K_{\alpha }(w|\tau )=\frac{e^{-\left(
p+1\right) \pi \frac{(Im w)^{2}}{Im \tau }}}{\eta (\tau )}\Theta \left[
\begin{array}{c}
\frac{\alpha }{4\left( p+1\right) } \\
0
\end{array}
\right] \left( 2\left( p+1\right) w|4\left( p+1\right) \tau \right) $, where
we introduce a non-analytic factor as in Ref. \cite{Cappelli}) and a neutral
component ($\chi _{\beta }$, the conformal blocks of the Ising Model).

The characters of the twisted sector are given by:
\begin{eqnarray}
\chi _{(0,s)}^{+}(w|\tau ) &=&\bar{\chi}_{\frac{1}{16}}\left( \chi
_{(0,s)}^{MR}(w|\tau )+\chi _{(2,s)}^{MR}(w|\tau )\right)  \label{ChtwTM} \\
\chi _{(1,s)}^{+}(w|\tau ) &=&\left( \bar{\chi}_{0}+\bar{\chi}_{\frac{1}{2}%
}\right) \chi _{(1,s)}^{MR}(w|\tau )
\end{eqnarray}
which do not depend on the parity of $p$;
\begin{eqnarray}
\chi _{(0,s)}^{-}(w|\tau ) &=&\bar{\chi}_{\frac{1}{16}}\left( \chi
_{(0,s)}^{MR}(w|\tau )-\chi _{(2,s)}^{MR}(w|\tau )\right) \\
\chi _{(1,s)}^{-}(w|\tau ) &=&\left( \bar{\chi}_{0}-\bar{\chi}_{\frac{1}{2}%
}\right) \chi _{(1,s)}^{MR}(w|\tau )
\end{eqnarray}
for $p$ even, and
\begin{eqnarray}
\chi _{(0,s)}^{-}(w|\tau ) &=&\bar{\chi}_{\frac{1}{16}}\left( \chi _{0}-\chi
_{\frac{1}{2}}\right) \left( K_{2s}\left( w|\tau \right) +K_{2(p+s)+2}\left(
w|\tau \right) \right) \\
\chi _{(1,s)}^{-}(w|\tau ) &=&\chi _{\frac{1}{16}}\left( \bar{\chi}_{0}-\bar{%
\chi}_{\frac{1}{2}}\right) \left( K_{2s+1}\left( w|\tau \right)
-K_{2(p+s)+3}\left( w|\tau \right) \right)
\end{eqnarray}
for $p$ odd. Notice that the last two characters are not present in the TM
partition function and that only the symmetric combinations $\chi
_{(i,s)}^{+}$ can be factorized in terms of the $c=\frac{3}{2}$ \ and $c=%
\frac{1}{2}$ theory. That is a consequence of the parity selection rule ($m$%
-ality), which gives a gluing condition for the charged and neutral
excitations. Furthermore the characters of the untwisted sector are given
by:
\begin{eqnarray}
\tilde{\chi}_{(0,s)}^{+}(w|\tau ) &=&\bar{\chi}_{0}\chi _{(0,s)}^{MR}(w|\tau
)+\bar{\chi}_{\frac{1}{2}}\chi _{(2,s)}^{MR}(w|\tau )=\chi
_{1,s}^{331}(w|\tau )  \label{vacuum1} \\
\tilde{\chi}_{(1,s)}^{+}(w|\tau ) &=&\bar{\chi}_{0}\chi _{(2,s)}^{MR}(w|\tau
)+\bar{\chi}_{\frac{1}{2}}\chi _{(0,s)}^{MR}(w|\tau )=\chi
_{2,s}^{331}(w|\tau ) \\
\tilde{\chi}_{(0,s)}^{-}(w|\tau ) &=&\bar{\chi}_{0}\chi _{(0,s)}^{MR}(w|\tau
)-\bar{\chi}_{\frac{1}{2}}\chi _{(2,s)}^{MR}(w|\tau )  \label{vacuum2} \\
\tilde{\chi}_{(1,s)}^{-}(w|\tau ) &=&\bar{\chi}_{0}\chi _{(2,s)}^{MR}(w|\tau
)-\bar{\chi}_{\frac{1}{2}}\chi _{(0,s)}^{MR}(w|\tau ) \\
\tilde{\chi}_{(s)}(w|\tau ) &=&\bar{\chi}_{\frac{1}{16}}\chi
_{(1,s)}^{MR}(w|\tau )=\chi _{3,s}^{331}(w|\tau )+\chi _{4,s}^{331}(w|\tau ),
\end{eqnarray}
where $\chi _{i,s}^{331}(w|\tau )$ are the characters of $331$ model \cite
{Halperin}:
\begin{eqnarray}
\chi _{1,s}^{331}(w|\tau ) &=&K^{0}\left( 0|\tau \right) K_{2s}\left( w|\tau
\right) +K^{2}\left( 0|\tau \right) K_{2(p+s)+2}\left( w|\tau \right)
\label{hal1} \\
\chi _{2,s}^{331}(w|\tau ) &=&K^{2}\left( 0|\tau \right) K_{2s}\left( w|\tau
\right) +K^{0}\left( 0|\tau \right) K_{2(p+s)+2}\left( w|\tau \right)
\label{hal2} \\
\chi _{3,s}^{331}(w|\tau ) &=&K^{1}\left( 0|\tau \right) K_{2s+1}\left(
w|\tau \right) +K^{3}\left( 0|\tau \right) K_{2(p+s)+3}\left( w|\tau \right)
\label{hal3} \\
\chi _{4,s}^{331}(w|\tau ) &=&K^{3}\left( 0|\tau \right) K_{2s+1}\left(
w|\tau \right) +K^{1}\left( 0|\tau \right) K_{2(p+s)+3}\left( w|\tau \right)
,  \label{hal4}
\end{eqnarray}
$K^{i}\left( 0|\tau \right) $ \ being the characters of the $c=1$ Dirac
theory. Notice that $K^{3}\left( -w|\tau \right) =K^{1}\left( w|\tau \right)
$, so that only for a balanced system the two characters can be identified
while $K^{0(2)}\left( -w|\tau \right) =K^{0(2)}\left( w|\tau \right) $. Let
us also point out that, as evidenced from Eq. (\ref{hal1}), one character of
the TM is identified with two characters of the\ $331$ model. In this way
the degeneracy of the ground state on the torus is reduced from $4\left(
p+1\right) $ to $3\left( p+1\right) $ when switching from $331$ to TM, a
clear signature of a transition from an Abelian statistics to a non-Abelian
one. Such a transition is due to the presence of two inequivalent Majorana
fermions together with the breaking of the symmetry which exchanges them. In
conclusion, while in the Halperin model the fundamental particles are Dirac
fermions with a well defined layer index, in the TM they are given in terms
of symmetric $\psi $ and antisymmetric $\bar{\psi}$ fields, that is as a
superposition of states belonging to different layers. As such, they behave
in a different way under twisted boundary conditions.

We point out that the partition function on the torus can be written as:
\begin{equation}
Z(\tau )=\frac{1}{2}\left( \sum_{s=0}^{p}2\left| \tilde{\chi}_{(s)}(0|\tau
)\right| ^{2}+Z_{untwist}^{+}(0|\tau )+Z_{untwist}^{-}(\tau
)+Z_{twist}^{+}(\tau )+Z_{twist}^{-}(\tau )\right)
\end{equation}
for $p$ even, where:
\begin{eqnarray}
Z_{untwist}^{+}(\tau )\text{{}} &=&\text{{}}\sum_{s=0}^{p}\left( \left|
\tilde{\chi}_{(0,s)}^{+}(0|\tau )\right| ^{2}+\left| \tilde{\chi}%
_{(1,s)}^{+}(0|\tau )\right| ^{2}\right) \\
Z_{untwist}^{-}(\tau )\text{{}} &=&\text{{}}\sum_{s=0}^{p}\left( \left|
\tilde{\chi}_{(0,s)}^{-}(0|\tau )\right| ^{2}+\left| \tilde{\chi}%
_{(1,s)}^{-}(0|\tau )\right| ^{2}\right)
\end{eqnarray}
\begin{eqnarray}
Z_{twist}^{+}(\tau )\text{{}} &=&\text{{}}\sum_{s=0}^{p}\left( \left| \chi
_{(0,s)}^{+}(0|\tau )\right| ^{2}+\left| \chi _{(1,s)}^{+}(0|\tau )\right|
^{2}\right) \\
Z_{twist}^{-}(\tau )\text{{}} &=&\text{{}}\sum_{s=0}^{p}\left( \left| \chi
_{(0,s)}^{-}(0|\tau )\right| ^{2}+\left| \chi _{(1,s)}^{-}(0|\tau )\right|
^{2}\right) ;
\end{eqnarray}
while for p odd we get simply:
\begin{equation}
Z(\tau )=\frac{1}{2}\left( \sum_{s=0}^{p}2\left| \tilde{\chi}_{(s)}(0|\tau
)\right| ^{2}+Z_{untwist}^{+}(0|\tau )+Z_{untwist}^{-}(\tau
)+Z_{twist}^{+}(\tau )\right) .
\end{equation}
As recalled above, the two Majorana fermions are not completely equivalent
and that reflects in the factorization of the partition function in the MR
and Ising (non-invariant) one:
\begin{equation}
Z(\tau )=Z_{MR}(\tau )Z_{\overline{I\sin g}}(\tau )
\end{equation}
where $Z_{MR}$ is the modular invariant partition function of the MR $c=3/2$
theory:
\begin{equation}
Z_{MR}(\tau )=\sum_{s=0}^{p}\left( \left| \chi _{(0,s)}^{MR}(0|\tau )\right|
^{2}+\left| \chi _{(1,s)}^{MR}(0|\tau )\right| ^{2}+\left| \chi
_{(2,s)}^{MR}(0|\tau )\right| ^{2}\right)
\end{equation}
and $Z_{\overline{I\sin g}}$ is the partition function of the Ising $c=1/2$
theory:
\begin{equation}
Z_{\overline{\text{I}\sin \text{g}}}(\tau )=\left| \bar{\chi}_{0}(\tau )%
\text{ }\right| ^{2}+\left| \bar{\chi}_{\frac{1}{2}}(\tau )\right|
^{2}+\left| \bar{\chi}_{\frac{1}{16}}(\tau )\right| ^{2}.
\end{equation}


\begin{thebibliography}{99}
\bibitem{cgm2}  G. Cristofano, G. Maiella, V. Marotta, \textit{Mod. Phys.
Lett. A}\textbf{\ 15} (2000) 1679.

\bibitem{AV1}  V. Marotta, A. Naddeo, \textit{J. High Energy Phys. }\textbf{%
08 }(2008) 029; V. Marotta, A. Naddeo, \textit{Nucl. Phys.\ B }\textbf{810}
(2009) 575.

\bibitem{prangegirvin}  R. E. Prange and S. M. Girvin (Eds.), \textit{The
Quantum Hall Effect}, 2nd Ed., Springer-Verlag, New York, 1990.

\bibitem{perspectives}  S. Das Sarma, A. Pinczuk (Eds.), \textit{%
Perspectives in Quantum Hall Effects}, Wiley, New York (1997).

\bibitem{dorda}  K. v.Klitzing, G. Dorda, M. Pepper, \textit{Phys. Rev. Lett.%
} \textbf{45} (1980) 494.

\bibitem{stormer}  D. C. Tsui, H. L. Stormer, A. C. Gossard, \textit{Phys.
Rev. Lett.} \textbf{48} (1982) 1559.

\bibitem{laughlin1}  R. B. Laughlin, \textit{Phys. Rev. B} \textbf{23}
(1981) 5632.

\bibitem{laughlin2}  R. B. Laughlin, \textit{Phys. Rev. Lett.} \textbf{50}
(1983) 1395.

\bibitem{ODLRO}  S.M. Girvin, A.H. MacDonald, \textit{Phys. Rev. Lett.}
\textbf{58} (1987) 1252.

\bibitem{jain}  J. K. Jain, \textit{Phys. Rev. Lett.} \textbf{65} (1989)
199; J. K. Jain, \textit{Phys. Rev. B} \textbf{40} (1989) 8079; J. K. Jain,
\textit{Phys. Rev. B} \textbf{41} (1990) 7653.

\bibitem{MR}  N. Read, G. Moore, \textit{Prog. Theor. Phys. Supp.} \textbf{%
107} (1992) 157.

\bibitem{FNTW}  E. Fradkin, C. Nayak, A. Tsvelik, F. Wilczek, \textit{Nucl.
Phys. B} \textbf{516} (1998) 704; M. Milovanovi\`{c}, N. Rezayi, \textit{%
Phys. Rev. B} \textbf{53} (1999) 13559.

\bibitem{fault1}  C. Nayak, S. H. Simon, A. Stern, M. Freedman, S. Das
Sarma, \textit{Rev. Mod. Phys. }\textbf{80} (2008) 1083.

\bibitem{teoria}  A. Stern, S. M. Girvin, A. H. MacDonald, N. Ma, \textit{%
Phys. Rev. Lett. }\textbf{86 }(2001) 1829; L. Balents, L. Radzihovsky,
\textit{Phys. Rev. Lett. }\textbf{86 }(2001) 1825; M. M. Fogler, F. Wilczek,
\textit{Phys. Rev. Lett. }\textbf{86 }(2001) 1833.

\bibitem{eisenstein}  J. P. Eisenstein, \textit{Solid State Comm.} \textbf{%
127} (2003) 123; J. P. Eisenstein, A. H. MacDonald, \textit{Nature} \textbf{%
432} (2004) 691.

\bibitem{Zee}  X. G. Wen, A. Zee, \textit{Phys. Rev. Lett. }\textbf{69 }%
(1992) 1811; X. G. Wen, A. Zee, \textit{Phys. Rev. B }\textbf{47 }(1993)
2265.

\bibitem{Zee1}  X. G. Wen, A. Zee, \textit{Int. J. Mod. Phys. B }\textbf{17\
}(2003) 4435.

\bibitem{girvin1}  S. M. Girvin, \textit{Int. J. Mod. Phys. B }\textbf{17\ }%
(2003) 4975.

\bibitem{YangMoon}  K. Moon, H. Mori, K. Yang, S. M. Girvin, A. H.
MacDonald, L. Zheng, D. Yoshioka, S. C. Zhang, \textit{Phys. Rev. B }\textbf{%
51 }(1995) 5138; K. Yang, K. Moon, L. Belkhir, H. Mori, S. M. Girvin, A. H.
MacDonald, L. Zheng, D. Yoshioka, \textit{Phys. Rev. B }\textbf{54 }(1996)
11644.

\bibitem{ncft1}  M. R. Douglas, N. A. Nekrasov, \textit{Rev. Mod. Phys. }%
\textbf{73} (2001) 977; A. Konechny, A. Schwarz, \textit{Phys. Rept. }%
\textbf{360} (2002) 353; R. J. Szabo, \textit{Phys. Rept. }\textbf{378}
(2003) 207.

\bibitem{landau1}  L. D. Landau, L. M. Lifschitz, \textit{Quantum Mechanics:
Non-Relativistic Theory}, Pergamon Press, Oxford (1977).

\bibitem{ncft3}  C. S. Chu, P. M. Ho, \textit{Nucl. Phys. B}\textbf{\ 550}
(1999) 151; M. M. Sheikh-Jabbari, \textit{Phys. Lett. B}\textbf{\ 455}
(1999) 129; D. Bigatti, L. Susskind, \textit{Phys. Rev. D}\textbf{\ 62}
(2000) 066004.

\bibitem{qhf1}  L. Susskind, \textit{hep-th}/0101029.

\bibitem{qh1}  R. B. Laughlin, ''Elementary theory of incompressible quantum
fluid'', in \textit{The Quantum Hall Effect}, R. E. Prange, S. M. Girvin
(Eds.), Springer, Heidelberg (1989).

\bibitem{top1}  Y. S. Wu, ''Topological Aspects of the Quantum Hall
Effect'', in \textit{Physics, Geometry and Topology}, H. C. Lee (Ed.),
Proceedings Banff (1989); R. Tao, Y. S. Wu, \textit{Phys. Rev.\ B} \textbf{30%
} (1984) 1097; D. Arovas, J. R. Schrieffer, F. Wilczek, \textit{Phys. Rev.
Lett.\ }\textbf{53} (1984) 722.

\bibitem{brane1}  T. Banks, W. Fischler, S. H. Shenker, L. Susskind, \textit{%
Phys. Rev. D }\textbf{55} (1997) 5112.

\bibitem{qhf2}  V. P. Nair, A. P. Polychronakos, \textit{Phys. Rev. Lett. }%
\textbf{87} (2001) 030403; D. Bak, K. M. Lee, J. H. Park, \textit{Phys. Rev.
Lett. }\textbf{87} (2001) 030402.

\bibitem{qhf3}  A. P. Polychronakos, \textit{J. High Energy Phys. }\textbf{%
11 }(2000) 008; S. Hellerman, M. van Raamsdonk, \textit{J. High Energy Phys.
}\textbf{10 }(2001) 039; A. Cappelli, M. Riccardi, \textit{J. Stat. Mech.:
Theor. Exper. }(2005) P05001; A. Cappelli, I. D. Rodriguez, \textit{J. High
Energy Phys. }\textbf{12 }(2006) 056.

\bibitem{qhf4}  A. P. Polychronakos, \textit{J. High Energy Phys. }\textbf{%
04 }(2001) 011; A. P. Polychronakos, \textit{J. High Energy Phys. }\textbf{%
06 }(2001) 070; B. Morariu, A. P. Polychronakos, \textit{Phys. Rev. D }%
\textbf{72} (2005) 125002.

\bibitem{alvarez1}  L. Alvarez-Gaum\`{e}, J. L. F. Barbon, \textit{Nucl.
Phys. B }\textbf{623 }(2002) 165.

\bibitem{uv1}  A. Das, M. M. Sheikh-Jabbari, \textit{J. High Energy Phys. }%
\textbf{06 }(2001) 028.

\bibitem{ncft2}  N. Seiberg, E. Witten, \textit{J. High Energy Phys. }%
\textbf{09 }(1999) 032.

\bibitem{morita}  A. Connes, \textit{Noncommutative Geometry}, Academic
Press, San Diego (1994); M. R. Douglas, \textit{hep-th}/9901146.

\bibitem{scw1}  A. Connes, M. R. Douglas, A. Schwarz, \textit{J. High Energy
Phys. }\textbf{02 }(1998) 003.

\bibitem{scw2}  A. Schwarz, \textit{Nucl. Phys. B }\textbf{534 }(1998) 720.

\bibitem{string1}  C. Hofman, E. Verlinde, \textit{J. High Energy Phys. }%
\textbf{12 }(1998) 010; C. Hofman, E. Verlinde, \textit{Nucl. Phys. B }%
\textbf{547 }(1999) 157; B. Pioline, A. Schwarz, \textit{J. High Energy
Phys. }\textbf{08 }(1999) 021.

\bibitem{wilson1}  J. Ambjorn, Y. M. Makeenko, J. Nishimura, R. J. Szabo,
\textit{J. High Energy Phys. }\textbf{05 }(2000) 023.

\bibitem{matrix1}  K. Saraikin, \textit{J. Exp. Theor. Phys.} \textbf{91}
(2000) 653; Z. Guralnik, J. Troost, \textit{J. High Energy Phys. }\textbf{05
}(2001) 022.

\bibitem{guralnik1}  Z. Guralnik, R. C. Helling, K. Landsteiner, E. Lopez,
\textit{J. High Energy Phys. }\textbf{05 }(2002) 025.

\bibitem{VM}  V. Marotta, \textit{J. Phys. A\ }\textbf{26} (1993) 3481; V.
Marotta, \textit{\ Mod. Phys. Lett.\ A}\textbf{\ 13} (1998) 853; V. Marotta,
\textit{\ Nucl. Phys.\ B }\textbf{527} (1998) 717; V. Marotta, A. Sciarrino,
\textit{Mod. Phys. Lett.\ A }\textbf{13} (1998) 2863.

\bibitem{cgm1}  G. Cristofano, G. Maiella, V. Marotta, \textit{Mod. Phys.
Lett. A }\textbf{15} (2000) 547.

\bibitem{cgm3}  G. Cristofano, V. Marotta, G. Niccoli, \textit{J. High
Energy Phys. }\textbf{06 }(2004) 056.

\bibitem{cgm4}  G. Cristofano, G. Maiella, V. Marotta, G. Niccoli, \textit{%
Nucl. Phys. B }\textbf{641 }(2002) 547.

\bibitem{noi1}  G. Cristofano, V. Marotta, A. Naddeo, \textit{Phys. Lett. B }%
\textbf{571} (2003) 250.

\bibitem{noi2}  G. Cristofano, V. Marotta, A. Naddeo, \textit{Nucl. Phys. B }%
\textbf{679 }(2004) 621.

\bibitem{ncmanybody}  A. P. Polychronakos, arXiv:0706.1095v2.

\bibitem{noi5}  G. Cristofano, V. Marotta, A. Naddeo, G. Niccoli, \textit{J.
Stat. Mech.: Theor. Exper. }(2006) L05002.

\bibitem{Halperin}  B. I. Halperin, \textit{Helv. Phys. Acta.\ } \textbf{56}
(1983) 1031; B. I. Halperin, \textit{Phys. Rev. B }\textbf{25\ }(1982) 2185.

\bibitem{matrix2}  D. B. Fairlie, P. Fletcher, C. K. Zachos, \textit{Phys.
Lett.\ B} \textbf{218} (1989) 203; D. B. Fairlie, C. K. Zachos, \textit{%
Phys. Lett.\ B} \textbf{224} (1989) 101; E. G. Floratos, \textit{Phys.
Lett.\ B} \textbf{228} (1989) 335.

\bibitem{membrane1}  M. Bordemann, E. Meinrenken, M. Schlichenmeier, \textit{%
Comm. Math.\ Phys.} \textbf{165} (1994) 281.

\bibitem{matrix3}  G. t'Hooft, \textit{Nucl. Phys.\ B} \textbf{138 }(1978)
1; G. t'Hooft, \textit{Nucl. Phys.\ B} \textbf{153 }(1979) 141.

\bibitem{AVfuture1}  V. Marotta, A. Naddeo, work in preparation.

\bibitem{kondo1}  I. Affleck, \textit{Acta Phys. Pol.} \textbf{26} (1995)
1869.

\bibitem{wen}  X. G. Wen, \textit{Int. J. Mod. Phys.\ B} \textbf{6 } (1992)
1711; X. G. Wen, \textit{Adv. in Phys.} \textbf{44} (1995) 405.

\bibitem{cristofano1}  G. Cristofano, G. Maiella, R. Musto, F. Nicodemi,
\textit{Nucl. Phys.\ B - Proceedings Supplements}\textbf{\ 33} (1993) 119.

\bibitem{top3}  E. Verlinde, \textit{Nucl. Phys. B }\textbf{300} (1988) 360.

\bibitem{AVfuture2}  V. Marotta, A. Naddeo, work in preparation.

\bibitem{Cappelli}  A. Cappelli, L. S. Georgiev, I. T. Todorov, \textit{%
Comm. Math. Phys. }\textbf{205\ }(1999) 657.

\bibitem{ho}  T. L. Ho, \textit{Phys. Rev. Lett.} \textbf{75} (1995) 1186.

\bibitem{gauge1}  A. Armoni, \textit{Nucl. Phys. B}\textbf{\ 593} (2001) 229.

\bibitem{ludaff1}  I. Affleck, A. W. W. Ludwig, \textit{Phys. Rev. Lett.}
\textbf{67} (1991) 161.

\bibitem{callan1}  C. G. Callan, I. R. Klebanov, J. M. Maldacena, A.
Yegulalp, \textit{Nucl. Phys.\ B} \textbf{443 }(1995) 444.

\bibitem{magneticbrane1}  J. Ambjorn, Y. M. Makeenko, G. W. Semenoff, R. J.
Szabo, \textit{J. High Energy Phys. }\textbf{02 }(2003) 026.

\bibitem{noi3}  G. Cristofano, V. Marotta, A. Naddeo, \textit{J. Stat.
Mech.: Theor. Exper. }(2005) P03006.

\bibitem{noi4}  G. Cristofano, V. Marotta, A. Naddeo, G. Niccoli, \textit{%
Eur. Phys. J. B}\textbf{\ 49 }(2006) 83.

\bibitem{noi6}  G. Cristofano, V. Marotta, A. Naddeo, G. Niccoli, \textit{%
Phys. Lett. A }\textbf{372} (2008) 2464; G. Cristofano, V. Marotta, A.
Naddeo, G. Niccoli, \textit{Phys. Lett. A }\textbf{372} (2008) 6965.

\bibitem{noi}  G. Cristofano, V. Marotta, P. Minnhagen, A. Naddeo, G.
Niccoli, \textit{J. Stat. Mech.: Theor. Exper. }(2006) P11009.

\bibitem{noi7}  G. Cristofano, V. Marotta, A. Naddeo, G. Niccoli, \textit{J.
Stat. Mech.: Theor. Exper. }(2008) P12010.

\bibitem{branehall2}  L. Cappiello, G. Cristofano, G. Maiella, V. Marotta,
\textit{Mod. Phys. Lett. A} \textbf{17} (2002) 1281.

\bibitem{branehall1}  B. A. Bernevig, J. H. Brodie, L. Susskind, N. Toumbas,
\textit{J. High Energy Phys. }\textbf{02 }(2001) 003.

\bibitem{tachyon1}  A. Sen, B. Zwiebach, \textit{J. High Energy Phys. }%
\textbf{03 }(2000) 002; K. Dasgupta, S. Mukhi, G. Rajesh, \textit{J. High
Energy Phys. }\textbf{06 }(2000) 022; J. A. Harvey, D. Kutasov, E. J.
Martinec, \textit{hep-th}/0003101.

\bibitem{AVfuture}  V. Marotta, A. Naddeo, work in preparation.
\end{thebibliography}
\end{document}